\documentclass[usegraphicx,usenatbib,useAMS]{mn2e}
\begin{document}

\title[Synthetic Stellar Populations]{Synthetic stellar populations: single
stellar populations, stellar interior models and primordial proto-galaxies}
\author[Jimenez et al.]
{Raul Jimenez$^1$, James MacDonald$^{2}$, James S. Dunlop$^{3}$,
Paolo Padoan$^4$, \and John A. Peacock$^3$ \\
$^1$Dept. of Physics and Astronomy, University of Pennsylvania, Philadelphia, PA 19104, USA.\\
$^2$Dept. of Physics and Astronomy, University of Delaware,
Newark, DE 19716,
USA.\\
$^3$Institute for Astronomy, University of Edinburgh, Royal
Observatory, Blackford Hill,
Edinburgh EH9-3HJ, UK. \\
$^4$Dept. of Physics, University of California, San Diego, CA
92093-0424, USA.}

\maketitle

\begin{abstract}
We present a new set of stellar interior and synthesis models for
predicting the integrated emission from stellar populations in star
clusters and galaxies of arbitrary age and metallicity. This work
differs from existing spectral synthesis codes in a number of
important ways, namely (1) the incorporation of new stellar
evolutionary tracks, with sufficient resolution in mass to sample
rapid stages of stellar evolution; (2) a physically consistent
treatment of evolution in the HR diagram, including the approach to
the main sequence and the effects of mass loss on the giant and
horizontal-branch phases. Unlike several existing models, ours yield
consistent ages when used to date a coeval stellar population from a
wide range of spectral features and colour indexes. We use {\em
Hipparcos} data to support the validity of our new evolutionary
tracks.
We rigorously discuss degeneracies in the age-metallicity plane and
show that inclusion of spectral features blueward of 4500 \AA\,
suffices to break any remaining degeneracy and that with moderate
$S/N$ spectra (10 per 20\AA\, resolution element) age and metallicity
{\em are not} degenerate. We also study sources of systematic errors
in deriving the age of a single stellar population and conclude that
they are not larger than $10-15$\%. We illustrate the use of single
stellar populations by predicting the colors of primordial
proto-galaxies and show that one can first find them and then deduce
the form of the IMF for the early generation of stars in the
universe. Finally, we provide accurate analytic fitting formulas for
ultra fast computation of colors of single stellar populations.
\end{abstract}


\section{Introduction}

The synthetic stellar spectrum of a galaxy is a well established
theoretical tool for investigating the properties of the integrated
light from distant galaxies where individual stars cannot be
resolved. Since the late 60's several groups have developed different
grids of synthetic stellar population models using a variety of
stellar interior tracks and observed or theoretical stellar spectra
(e.g.
\citet{Tinsley_68,BB77,Renzini_81,Bruzual_83,BO86,Arimoto_Yoshii_87,GR_87,BC93,Worthey_94,BCF94,FiocRocca97,JPMH98,LLG02}). The
idea is very simple: stars are born with a given initial mass function
(IMF) and they evolve in time according to stellar evolution. At any
time one can compute the integrated spectrum by summing up the
individual spectra of the stars in the population at that instant. Two
major approaches have been used to compute the integrated light of a
stellar population: the fuel consumption theorem
\citep{Renzini_81,Renzini_Buzzoni_83} and the isochrone technique,
 first developed by \citet{BB77} and later used by
\citet{Charlot_Bruzual_91}. The fuel consumption theorem simply uses
the fact that the contribution of stars in any post main--sequence
evolutionary stage is proportional to the amount of nuclear fuel they
burn at that stage, and approximates the post main--sequence evolution
of the stars in the integrated population by the stellar track of the
most massive star alive at the main sequence turn--off (MSTO). In
contrast, the isochrone technique uses a continuous distribution of
stellar masses, and hence tracks, to compute the locus in luminosity
and $T_{\rm eff}$ at a given time for any mass. From this, a smooth
isochrone can be computed. The fuel consumption theorem remains an
elegant method of studying the fastest stages in the evolution of any
population. On the other hand, with the new generation of fast
computers, the isochrone technique is clearly the most accurate and
precise method of computing the integrated light of any stellar
population.

Regardless of the computational technique used to calculate the integrated
spectrum of a stellar population, the most important ingredient remains the
stellar input: both stellar interior and stellar atmospheric models.
Disagreement over stellar interior models (convection, mass loss, opacities),
the modelling of post main sequence evolutionary stages and the modelling of
stellar atmospheres (opacities, NLTE effects, bolometric correction, mass
loss, etc.), combine to make the derived age of even a simple (i.e., no dust,
no AGN contamination) stellar population vary by about 10\% (for a fixed mass
and metallicity), depending on which of the currently available synthetic
stellar population codes (e.g. \citet{Charlot_Worthey_Bressan_96,Spinrad+97})
is used to interpret the data (see section 4). A similar problem occurs when
trying to date Globular Clusters in the Galaxy, the age of which is currently
uncertain by about 10\% \citep{Chaboyer95,Jimenez+96,ChaboyerKrauss02}.

We have been motivated to attempt to improve this situation by the
fact that the new generation of large 8-10m telescopes can now deliver
spectra of galaxies at $z>1$ of sufficient quality to merit accurate
age dating. The accurate determination of the ages of high-redshift
galaxies can yield important constraints not only on models of galaxy
formation, but also on the age of the universe. In particular, in a
series of papers \citep{D+96,Spinrad+97,NDJ01,NDJH03} we have
addressed the issue of determining the ages of the reddest known
elliptical galaxies at $z \geq 1$. To aid in the interpretation of our
data and others, we have developed a set of simple synthetic stellar
population models (SSP) that overcomes some of the problems described
above.

Most of the disagreement between the existing synthetic stellar evolution
codes stems from the difficulty of modelling accurately the post main-sequence
evolution, both because the physics involved in these stages is not completely
known (opacities, convection, nuclear rates) and because mass loss strongly
controls the final fate of the evolution of the star in these phases. Ideally,
one would like to have a robust set of stellar models that are computed
self-consistently (i.e. interior, photosphere and chromosphere computed at the
same time) and that include the effects of mass loss and dust grain
formation. However this is not yet possible, and in any case it is important
to realise that mass loss {\em cannot} be incorporated as a {\em fixed
  quantity} for all stars in the population since it varies from star to star.

In the new synthetic stellar population models presented here we have
endeavoured to improve the ability of the modelling to accurately reproduce the
post-main sequence evolution of real stellar populations by incorporating an
algorithm which has been previously applied with success to a number of other
stellar evolution problems (e.g.
\citet{Jorgensen_91,JT93,Jimenez+96,Jorgensen_Jimenez_97}). This algorithm
accurately simulates the evolution of all post main-sequence evolution stages,
and includes a proper modelling of the HB along with an accurate account of
the formation of carbon stars on the AGB. Furthermore, the mixing length
parameter and the mass loss are properly calibrated using the position of the
RGB and the morphology of the HB in real star clusters, respectively.

The other main new feature of our spectral synthesis models is the
incorporation of new stellar evolutionary tracks, with sufficient resolution
in mass to sample rapid stages of stellar evolution. As a result of these
improvements (which we describe in detail in this paper), unlike several
existing population codes, our models yield consistent ages when used to date
a coeval population from a wide range of spectral features and colour indexes.

Our new SSP code has been applied successfully to a variety of different
populations. it has been used to determine the ages of high redshift galaxies
\citep{D+96}, the ages of Low Surface Brightness Galaxies
\citep{PJA97,JPMH98}, the role of star formation and the Tully-Fisher law
\citep{HJ99} and the age of the Galactic disc \citep{JFK98}. The purpose of
this paper is to present the new library of synthetic stellar population
spectra and discuss in more detail the physics and assumptions in our SSP
modelling procedure.

The paper is organised as follows: in \S 2 we present the new set of
stellar interior tracks and discuss their accuracy when confronted
with individual stellar observations. The synthesis models are
presented in \S 3. The degeneracies in the age-metallicity plane are
discussed in \S 4 while systematic errors are considered in \S 5. The
application of synthesis models to primordial proto--galaxies is
presented in \S 6 along with a method to determine the initial mass
function of these galaxies. \S 7 discusses the IR flux density and
detectability of primordial proto--galaxies. Our conclusions are
presented in \S 8. An appendix provides fitting formulas for computing
broad band colors of SSPs.

\section{Stellar models and physics input}

\subsection{Library of stellar interiors}

We have computed a new interior stellar library with the stellar
evolution code JMSTAR developed by one of us (JM) from the code of
\citet{Eggleton71,Eggleton72}.

\begin{figure*} \includegraphics[width=17cm,height=18cm]{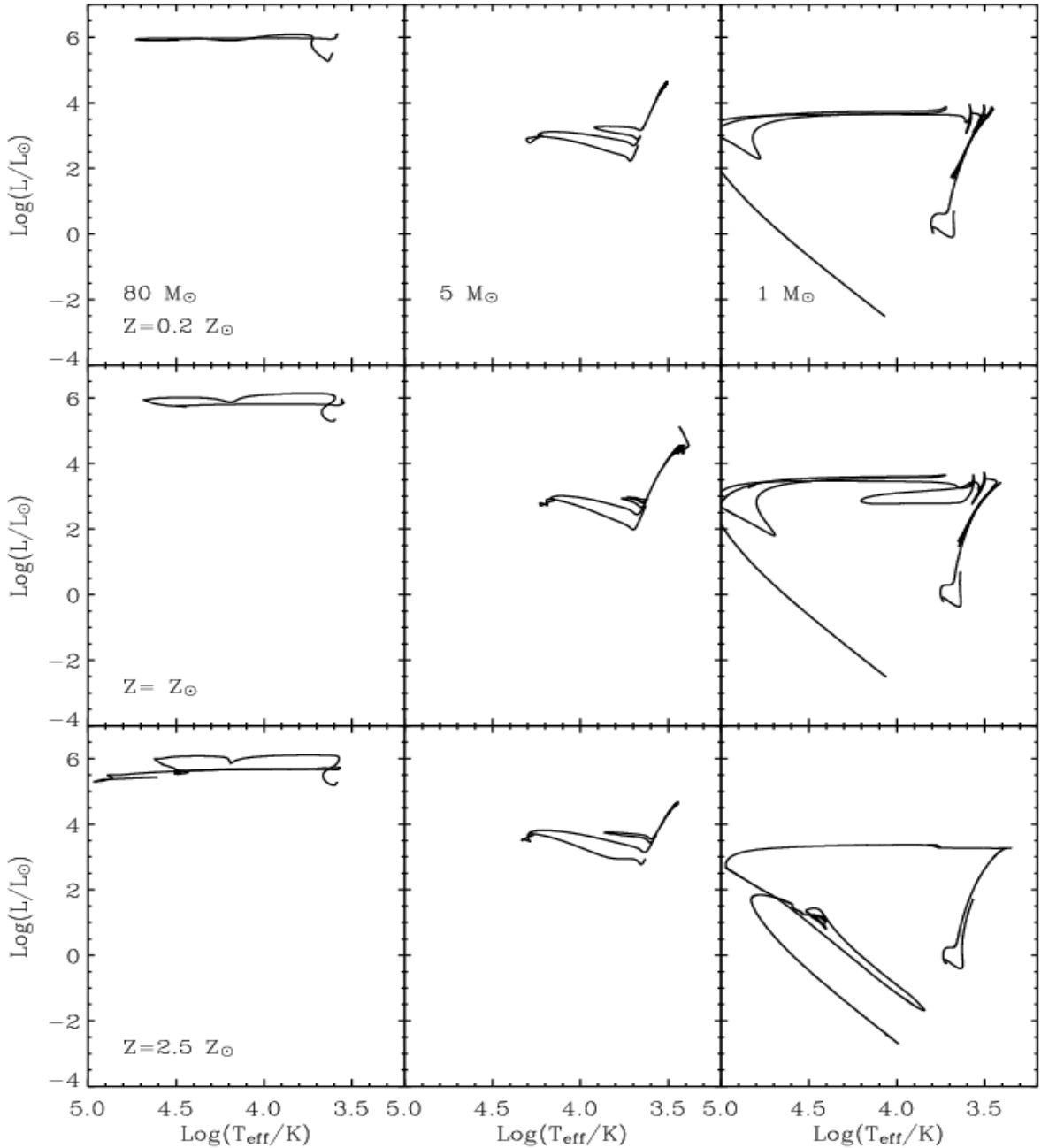}
\caption{Stellar tracks in the T${\rm eff}-$L plane computed with
JMSTAR are shown for 3 masses and metallicities. The tracks have been
evolved from the initial Hayashi contracting phase until the white
dwarf phase or carbon ignition, depending on the mass. The tracks are
evolved in one run without interruptions and are able to go through
the different He flashes in the AGB phase (see text for more
details).} \label{fig:plotiso} \end{figure*}

In this code, the whole star is evolved by a relaxation method without
use of separate envelope calculations. Some of the advantages of this
approach are that (1) gravothermal energy generation terms are
automatically included for the stellar envelope, (2) mass loss occurs
at the stellar surface rather than an interior point of the star, and
(3) the occurrence of convective dredge-up of elements produced by
nucleosynthesis in the interior to the photosphere is clearly
identifiable. The code uses an adaptive mesh technique similar to that
of \citet{WNN84}. Advection terms are approximated by second-order
upwind finite differences. Convective energy transport is modeled by
using the mixing-length theory described by
\citet{Mihalas_78}. Composition equations for H, $^3$He, $^4$He,
$^{12}$C, $^{14}$N, $^{16}$O, and $^{24}$Mg are solved simultaneously
with the structure equations. Composition changes due to convective
mixing are modeled in the same way as \citet{Eggleton72} by adding
diffusion terms to the composition equations. However, the
prescription for the diffusion coefficient differs from that of
\citet{Eggleton72}. The diffusion coefficient is consistent with
mixing-length theory \citep{IbenMacDonald95}, $\sigma_{\rm con} =
\beta w_{\rm con} l$, where $w_{\rm con}$ is the convective velocity,
$l$ is the mixing length, and $\beta$ is a dimensionless convective
mixing efficiency parameter. OPAL radiative opacities
\citep{Iglesias_Rogers_96} are used for temperatures (in kelvins)
above $\log_{10}T = 3.84$. For temperature below $\log_{10} T =
3.78$, we use opacities kindly supplied by D.  Alexander and
calculated by the method of \citet{Alexander_Ferguson_94}. Between
these temperature limits, we interpolate between the OPAL and
Alexander opacities. Nuclear reaction rates are taken from
\citet{Angulo+99} with screening corrections from
\citet{Salpeter_vanHorn69} and \citet{Itoh+79}. Neutrino loss rates
are from \citet{BPS67} with modifications for neutral currents
\citep{Ramadurai76}. Plasma neutrino rates are from \citet{HRW94}. The
equation of state is determined by minimization of a model free energy
(see, e.g., \citet{FGV77}) that includes contributions from internal
states of the $H_2$ molecule and all the ionization states of H, He,
C, N, O, and Mg. Electron degeneracy is included by the method of
\citet{EFF73}. Coulomb and quantum corrections to the equation of
state follow the prescription of \citet{IFM92}, with the Coulomb free
energy updated to use the results of \citet{SDS90}. Pressure
ionization is included in a thermodynamically consistent manner by use
of a hard-sphere free-energy term.

Mass loss is included by using a scaled \citet{Reimers75} mass-loss law,

\begin{equation}
\dot M=\eta_R 1.27 \times 10^{-5} M^{-1} L^{1.5} T_{\rm eff}^{-2}
M_{\odot} yr^{-1} \end{equation}

for cool stars (T$_{\rm eff} \le 10^4$ K) and an approximation to the
theoretical result of \citet{Abbott82},

\begin{equation}
\dot M=-1.2 \times 10^{-15} \frac{Z}{Z_{\odot}} \left (
\frac{L}{L_{\odot}} \right )^2 \left ( \frac{M_{\rm eff}}{M_{\odot}}
\right )^{-1} M_{\odot} {\rm yr}^{-1} \end{equation}

for hot stars ($T_{\rm eff} \ge 10^4$ K). For AGB stars we also
include additional mass loss at a rate obtained by fitting the
observed mass loss rates for Mira variables \citep{Knapp+98}.

\begin{equation}
\dot M_{AGB}= -3.85 \times 10^{-12} \left ( \frac{M}{M_{\odot}} \right
)^{-4.61} \left ( \frac{R}{100 R_{\odot}} \right )^{11.76} M_{\odot}
yr^{-1} \end{equation}

The surface boundary condition is treated as follows. We assume that
the atmosphere is plane-parallel and thin. A small value of optical
depth (typically $\tau = 0.01$) is assigned to the center of the
outermost zone of the stellar model. The corresponding surface
temperature $T_s$ is related to the effective temperature, $T_{\rm
eff}$, by the Eddington approximation $T^4_s=T^4_{\rm eff} (0.5+0.75
\tau)$. The surface gas pressure satisfies $p_gs= \tau (g_es/k)$ where
$g_es$ is the effective gravity (surface gravity reduced by the
effects of radiation pressure) at the surface of the star.  We stress
that the code treats the surface boundary on an equal footing with all
other shells in the star.

\begin{figure} \includegraphics[width=8cm,height=6cm]{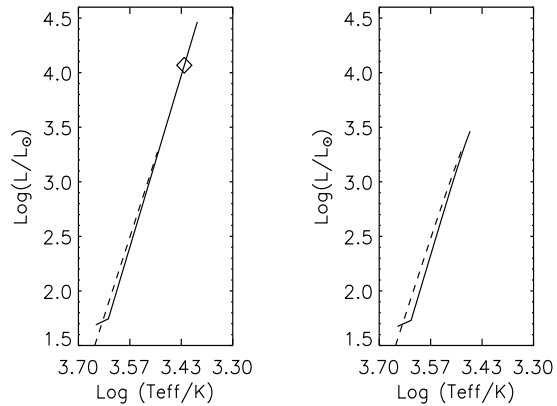}
\caption{Evolution of a solar mass star with solar metallicity under
the assumption that no mass loss occurs (left panel) or that it does
($\eta=0.4$; right panel). The dashed line is the red giant branch
while the solid line is the asymptotic giant branch. The difference in
luminosity, for the termination of the AGB, between the two cases is
quite apparent (almost an order of magnitude). The diamond shows when
a carbon star will be produced. By not including mass loss in stellar
tracks one can seriously overestimate the luminosity of the giants in
stellar populations.} \label{fig:rgbsynt} \end{figure}

We have used the evolution code to create a grid of interior models
with the following parameters. We computed stellar tracks for five
different (solar scaled) metallicities: $Z=0.0002, 0.004, 0.01, 0.02$
and $0.05$. We fixed the helium content for the solar model ($Y=0.27$)
as to reproduce the luminosity of the Sun at the present age and
adopted $dY/dZ=2$, consistent with the findings of \citet{JFMG03}.  We
show some representative stellar tracks in fig~\ref{fig:plotiso} for
different metallicities. The tracks were evolved from an initial
Hayashi phase to the very late stages in a {\em continuous} run. Thus
it was possible to deal with the different flashes along the evolution
during the giant phases. The evolution of low mass stars ($M \le 3
M_{\odot}$) is ended after they settle onto the white dwarf cooling
sequence. The evolution of massive stars ($M \ge 10 M_{\odot}$) is
stopped at the end of carbon burning or if the central temperature
exceeds $10^9 K$. For the intermediate mass stars, the calculation is
stopped once the envelope begins to be ejected hydrodynamically due to
the radiative luminosity exceeding the local Eddington limit.

\begin{table}
\begin{center}
\begin{tabular}{c|c}
\hline
$Z$ & $Y$ \\
\hline
0.0002 &  0.235   \\
0.004  &  0.24    \\
0.01   &  0.25    \\
0.02   & 0.27     \\
0.05   & 0.33     \\
\hline
\end{tabular}
\caption{Metallicity and helium fractions for the grid of stellar interior
models computed with JMSTAR and used to assemble the grid of synthetic
stellar population models. All stellar tracks were computed using
solar-scaled abundances.}
\end{center}
\end{table}

\begin{figure} \includegraphics[width=8cm,height=8cm]{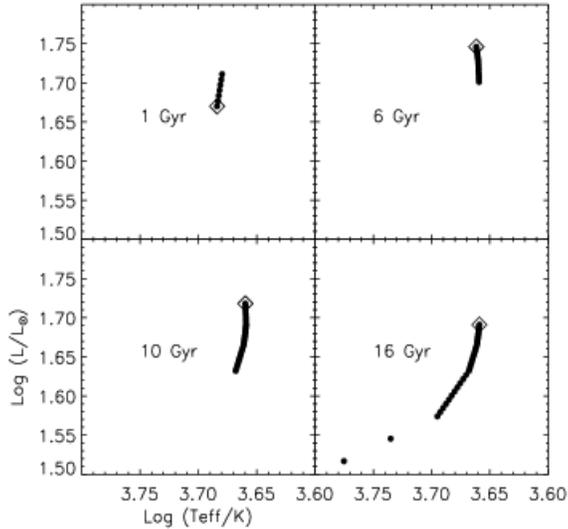}
\caption{The morphology of the HB is shown for 4 different ages (all
for solar metallicity). We have use a gaussian distribution for the
mass loss parameter ($\eta$) around a central value of $0.4$. Without
this the HB would be a point. Note that for ages less than 10 Gyr the
HB is actually vertical.} \label{fig:hbmorph} \end{figure}

In order to produce a model for the synthetic stellar population it is
necessary to have a set of stellar interior tracks that includes {\em
all} stellar evolutionary stages for a large range of masses. It is
well known that the main discrepancies between different stellar
population codes arise from the different treatment of the late stages
of stellar evolution \citep{Charlot_Worthey_Bressan_96}. Since a
proper and accurate modelling of the HB and AGB is extremely important
for predicting the UV (HB) and IR (AGB) properties of any stellar
population, we have included in our SSP code an accurate
semi-empirical algorithm for computing the evolution of the HB and
AGB. This is very important since evolution after the RGB cannot be
described on the basis of one choice of mass loss parameter (this
would result in the absence of a HB in globular clusters). The
analytical description of the effect of mass loss (and the influence
of different choices of mixing length) in our semi-empirical
algorithm, makes it particular useful in handling realistic population
descriptions where variation from star to star in the mass loss
efficiency needs to be described as a distribution function.  We
therefore believe that we can obtain the highest accuracy by use of a
semi-empirical algorithm. A direct comparison between the
semi-empirical method and tracks computed with JMSTAR can be found in
\citet{Jimenez+95}.  The fast and accurate computation of the HB, AGB
and TP-AGB, allowed by the semi-empirical method makes it possible to
produce stellar tracks for several values of the mass loss parameter
($\eta$ in the Reimers formulation, or any other similar) and
mixing-length parameter ($\alpha$). Therefore, it is possible to
analyse the impact of these (unknown) parameters in the final stellar
track (see Fig.~\ref{fig:rgbsynt}). In two previous studies
\citep{JT93,Jimenez+96} we demonstrated that the morphology of the HB
is due to star-to-star variations in the mass loss parameter. This
automatically gives a powerful tool to model the HB including, a
priori, a realistic distribution of values for the mass loss parameter
in the RGB.

Mass loss takes place at the very end of the RGB and is most often
described by the empirical Reimers mass loss law \cite{Reimers75}
which is derived from measurements of M type giants. However, instead
of following the evolution along the RGB using a fixed value for the
$\eta$ parameter (as is usually done in the literature, and as was
done by Reimers himself) we proceed in a different way. We use the
computed grid of stellar interior models and calculate interpolation
formulas (for the RGB) for $L$, $T_{\rm eff}$, $M$, $M_c$ (core mass),
$Z$, $Y$ and $\alpha$ (mixing length). The fitting formulas can be
found in \citet{Jimenez+96} and for brevity we will not repeat them
here. The average value of $\eta$ is determined by matching the
observed HB mass distribution. A very fast and accurate numerical
computation of the evolution along the RGB can then be performed by
taking advantage of the fitting formulas. The addition of mass to the
core during a given time step in the integration along the RGB is
determined on the basis of the instantaneous luminosity, the known
energy generation rate, and the length of the time step. The mass of
the core ($M_c$) at the end of the time step determines the new value
of $L$ according to the fitting formulas. The total stellar mass at
the end of each time step is calculated as the mass at the beginning
of the time step minus the mass loss rate times the length of the time
step. The evolution of the synthetic RGB track is stopped when $M_c$
reaches the value determined in the fitting formulas for the He-core
flash. A more detailed study of this approach and its comparison with
other stellar models can be found in \citet{Jimenez+95}.

Evolution along the AGB is much faster than the corresponding
evolution in the RGB and HB, lasting around $5 \times 10^6$ years. It
has two differentiated phases: the early AGB (E-AGB) and the
TP-AGB. The E-AGB occurs after the helium core is completely converted
to oxygen (and smaller amounts of carbon) and starts when the star is
burning helium in a hydrogen-exhausted shell. This phase ends when the
hydrogen shell has passed through a minimum in luminosity. The TP-AGB
consists of a phase of double-shell burning with thermal-shell flashes
and strong mass loss. The modelling of the AGB evolution is rather
complicated due to the above considerations. We have modelled it using
the prescriptions in \citet{Jorgensen_91} and will not repeat them
here. The most important issue is that we can compute when (or not)
the star along its AGB will become carbon rich as well as whether
the star will end the AGB as a planetary nebula or not. The advantage
of this approach to modelling the AGB is that, as for the HB, we can
produce accurate predictions on the fate of the star and explore a
large range of parameters for the mass loss, mixing length, chemical
composition, etc, in a fast and accurate way. The interpolation
formulas are based on numerical AGB computations from JMSTAR, as
described in \citet{Jorgensen_91}, and the evolution is terminated
when the core mass reaches the total mass of the star, or the total
mass minus the mass of a planetary nebula.

\begin{figure} \includegraphics[width=8cm,height=6cm]{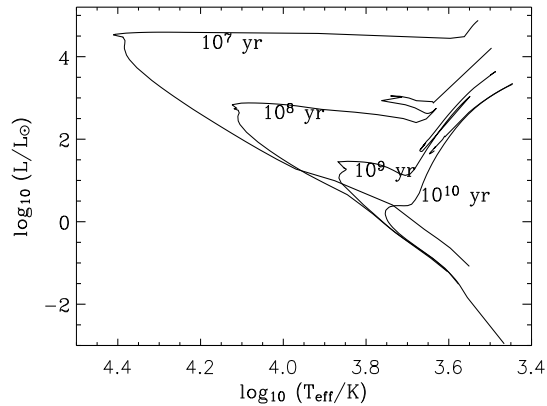}
\caption{Four isochrones are shown for the solar metallicity case. The
post AGB phase is not shown for clarity. Note that at $10^7$ years the
low mass stars have not reached the main sequence yet.}
\label{fig:fewiso} \end{figure}

To summarize, the most important advantages of our method for
isochrone synthesis are as follows:

\begin{itemize}

\item A state-of-the art code to generate stellar interiors that can
evolve the stellar track from the initial contracting gaseous phase to
the final stages without interruption and that includes the latest
advances in reaction rates and opacities.

\item A realistic distribution function for the mass loss efficiency
parameter
  $\eta$ is determined on the basis of the HB morphology. The actual
values of
  $\eta$ used in the present work are taken from our previous studies of
  globular clusters \cite{Jimenez+96}. In particular, this distribution
  ensures a realistic integrated value of the blue and UV radiation from
the
  SSP.

\item The mixing-length parameter is determined from fits to globular
clusters
  with known metallicities. This is very important for a realistic
description
  of the red and IR radiation from the SSP models.

\item The semi-empirical algorithm has the same accuracy as the original
  numerical computations on which it is based (plus being improved in
$\alpha$
  and $\eta$ based on observations), but it is computationally much
faster.
\end{itemize}

In order to illustrate the effect that changes in mass loss can have,
we have computed, using the above prescription, the evolution of the
Sun ($Z=0.02, Y=0.28$) with and without mass loss (using the Reimers
formalism with $\eta=0.4$). In the first case the star evolves well
into the AGB and develops into a carbon star before it will eventually
transform through the post-AGB phase into a WD. In the second case the
star leaves the AGB at a much lower luminosity (10 times lower) and at
a greater $T_{\rm eff}$ (3000K instead of 2500K). The consequence for
synthetic stellar population models are quite obvious: if mass loss is
ignored the red-IR luminosity of the population is overestimated (see
Fig.~\ref{fig:rgbsynt}).

Stellar isochrones are computed by interpolating stellar tracks at
equivalent stellar evolutionary points as defined by
\citet{SSMM92}. This is important in order to have a good sampling of
all stellar evolutionary stages. Some examples of stellar isochrones
are shown in Fig.~\ref{fig:fewiso}, where for clarity we are not
plotting the post-AGB phase.

\subsection{Library of stellar photospheres}

To compute the synthetic spectra corresponding to our stellar
evolutionary interior models, we have used a set of theoretical
stellar photospheres for single stars, most of them come from the most
recent Kurucz library ({\tt
http://kurucz.harvard.edu/grids.html}). The accuracy of the Kurucz
models has been the subject of recent studies. They do an excellent
job in reproducing broad band colors of individual stars
\citep{Bessell98} stars. Some comparison has been done also between
the Kurucz models and the, low resolution, spectra of about solar
metallicity star \citep{W94} with equally good results. The models are
known to not fare that well for wavelengths below $\sim 2500$ \AA\, or
for very low temperatures. For temperatures cooler than $4000K$ we
have generated our own (LTE) models using the the version of the MARCS
code developed by Uffe Jorgensen (private communication).  The above
library of atmospheres is used to compute spectra and magnitudes
(using the appropriate filters) for the stellar interior
isochrones. We choose theoretical spectral libraries as for them one
knows {\em exactely} the value of $T_{\rm eff}$, $log g$ and the
metallicity, thus one can match exactely them to the interior
library. This is not the case for observed spectra, for which the
above parameters are not exactely known, and worse, none of them have
exactely the same value as the one needed for the interior models. We
therefore much prefer to build theoretical models (contrast them with
individual stars) and use them to predict the spectral energy
distribution from the corresponding interior model.

\subsection{Validity of our stellar interior models}

Here we show that our stellar models fit observations of {\em
individual stars}. The first test concerns stars of different
metallicities in the solar neighborhood with {\em accurate distances},
while the second test benchmarks the models against star clusters in the
Milky Way.

\begin{figure}
\includegraphics[width=8cm,height=8cm]{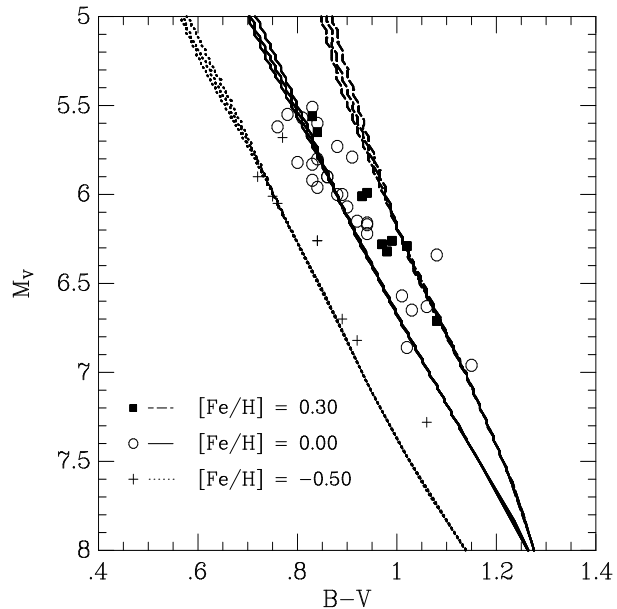}
\caption{Comparison of some of our isochrones with appropriate stars
from the Hipparcos catalog. The isochrones are from the left
[Fe/H]=0.50, 0.00 and 0.30 and shown by dotted, solid and dashed,
respectively. Crosses refer to the stars with the same metallicity as
the metal-weak isochrones, open circles are the stars matching with
the solar isochrones and squares are the stars matching with the
metal-rich isochrones. The metal-weak and solar isochrones fit the
data very well, while the metal-rich isochrone provides a less
satisfactory fit for $M_{V} < 6$. The ages of the isochrones are 10,
12 and 14 Gyr.}
\label{fig:hippcomp}
\end{figure}

\begin{figure*}
\includegraphics[width=18cm,height=12cm]{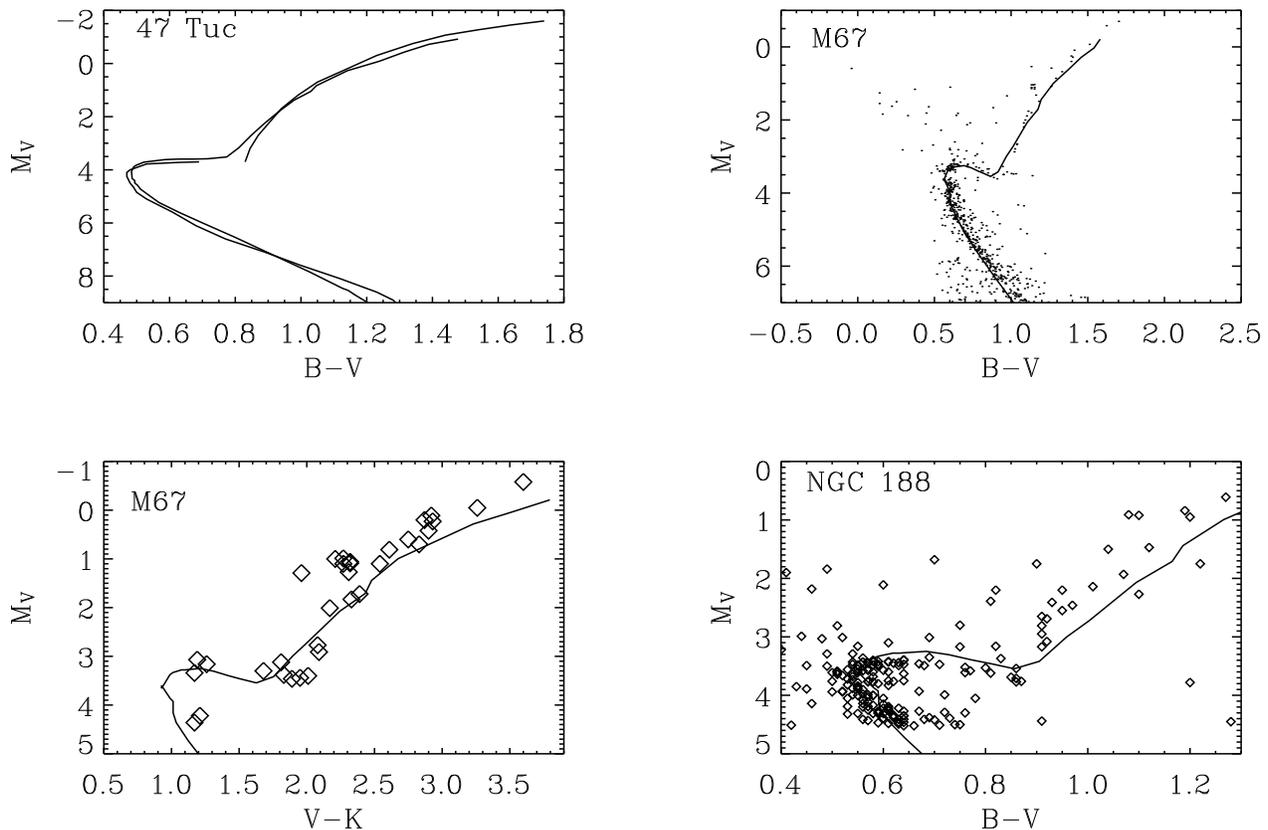}
\caption{Observed colour magnitude diagram for three stellar
clusters. The solid line corresponds to the best fitting isochrone
from our models. For 47Tuc the broken solid line is a fit to the
observations. Note the good fit to the overall shape of the
color-magnitude diagram, both the MSTO and giant branches. See
text for more details.} \label{fig:clusters}
\end{figure*}

We have used the sample of stars from the Hipparcos catalog described
in \citet{KFJ02} and for which accurate metallicities have been
derived either from spectroscopy or narrow band photometry. The sample
consists of 213 stars with metallicities ranging from $-1.3 < $ [Fe/H]
$ < 0.3$. All stars have accurate photometry and parallax distances
from Hipparcos (see \citet{KFJ02}). In Fig.~\ref{fig:hippcomp} we show
the locus of a set of isochrones from our stellar library for
different metallicities and ages (at such low luminosities the age
makes no difference in the locus of the isochrone). The different
symbols correspond to Hipparcos stars from the above mentioned catalog
and with the appropriate metallicity. Both solar and sub-solar models
provide a good fit to the data, while the super solar model is not so
good at higher luminosities but provides a fair fit to the data for
$M_V < 6$.  \citet{KFJ02} discuss how other independent sets of
isochrones fit the data and conclude that our isochrones are among the
best fit to the Hipparcos data. Despite this, attention should be
paid to super-solar models since they seem to be difficult to
construct.

Having demonstrated that the isochrones provide a good fit to dwarfs
of different metallicities we explore how successful they are at
fitting the giant branches. Since the age does play a role in the
temperature of the giant branch, the test with Hipparcos local stars
is not possible. In order to attempt this test we are therefore forced
to use star clusters which have much more uncertain distances (and
therefore less certain ages as well) than Hipparcos nearby stars but
have a simpler star formation history: they are single bursts. However
the colour of the giant branches is well determined and therefore
provides a good test for the models, specially at low temperatures. So
we do adopt distances published in the literature for these cluster
and fit the best isochrone to the overall color-magnitude diagram.

In Fig.~\ref{fig:clusters} we show fits to the color-magnitude diagram
of several globular and open clusters. In all cases the isochrones
provide good fits to the overall shape of the color-magnitude
diagram. Further, both the red giant branches and main sequence turn
off are well fitted by the isochrones. The fit is good for different
colours: the red giant branch of M67 is well fitted for $B-V$ and
$V-K$ as well. The fit to the globular cluster 47 Tuc has been done for
an age of 11 Gyr and $[Fe/H]=-0.76$. For M76 we used $[Fe/H]=-0.1$ and
an age of 5 Gyr. Finally, NGC188 has been fitted with an isochrone
with $[Fe/H]=-0.1$ and 6 Gyr.

We conclude that the stellar interior models provide good fits to
available photometry of single stellar populations and individual
stars. We are most encouraged by the good fit the isochrones provide to
Hipparcos dwarfs since these are the only stars with accurate
distances, and therefore absolute magnitude, and metallicities and
their color magnitude diagram does not depend on the age of the
star. It is also reassuring that the colours of theoretical red giant
branches are able to reproduce those of observed clusters.

\begin{figure*}
\includegraphics[width=18cm,height=10cm]{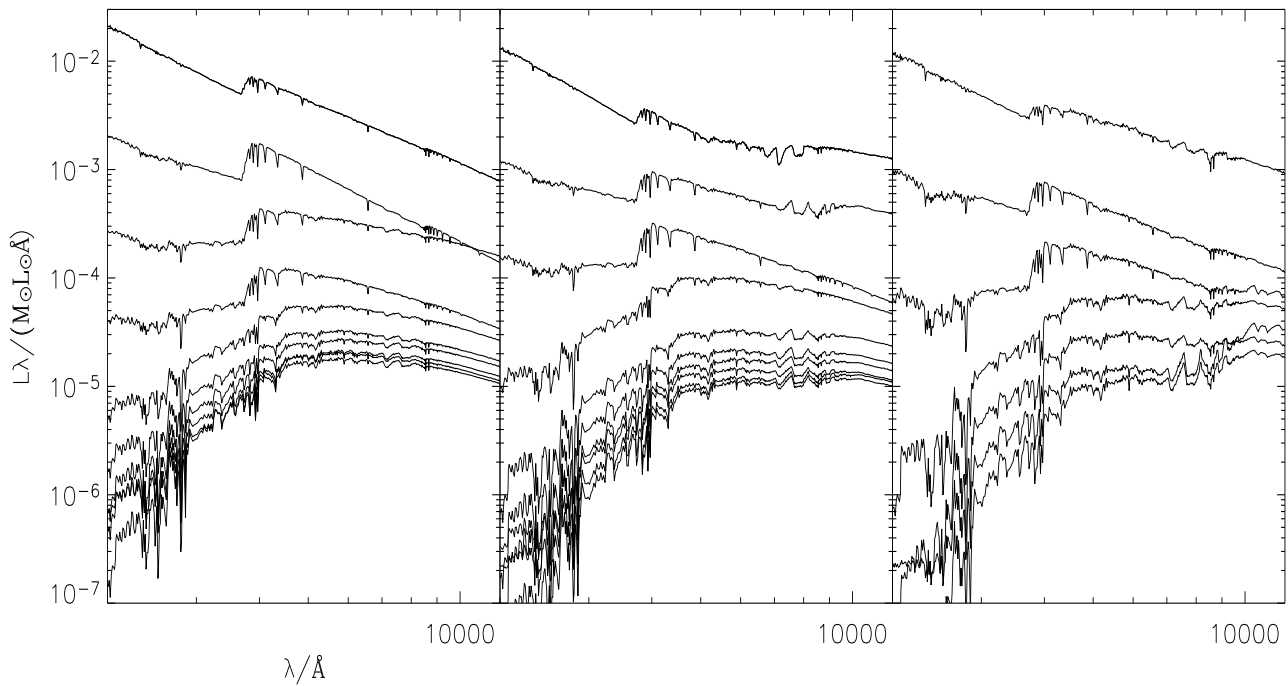}
\caption{Spectral energy distributions of single stellar populations
for three different metallicities: $0.2, 1$ and twice solar (from left
to right) and ages between $10^7$ and $13 \times 10^{9}$ years (from
top to bottom in each panel).}
\label{fig:fewssp}
\end{figure*}

\section{The library of synthetic stellar population spectra}

\begin{figure}
\includegraphics[width=8cm,height=8cm]{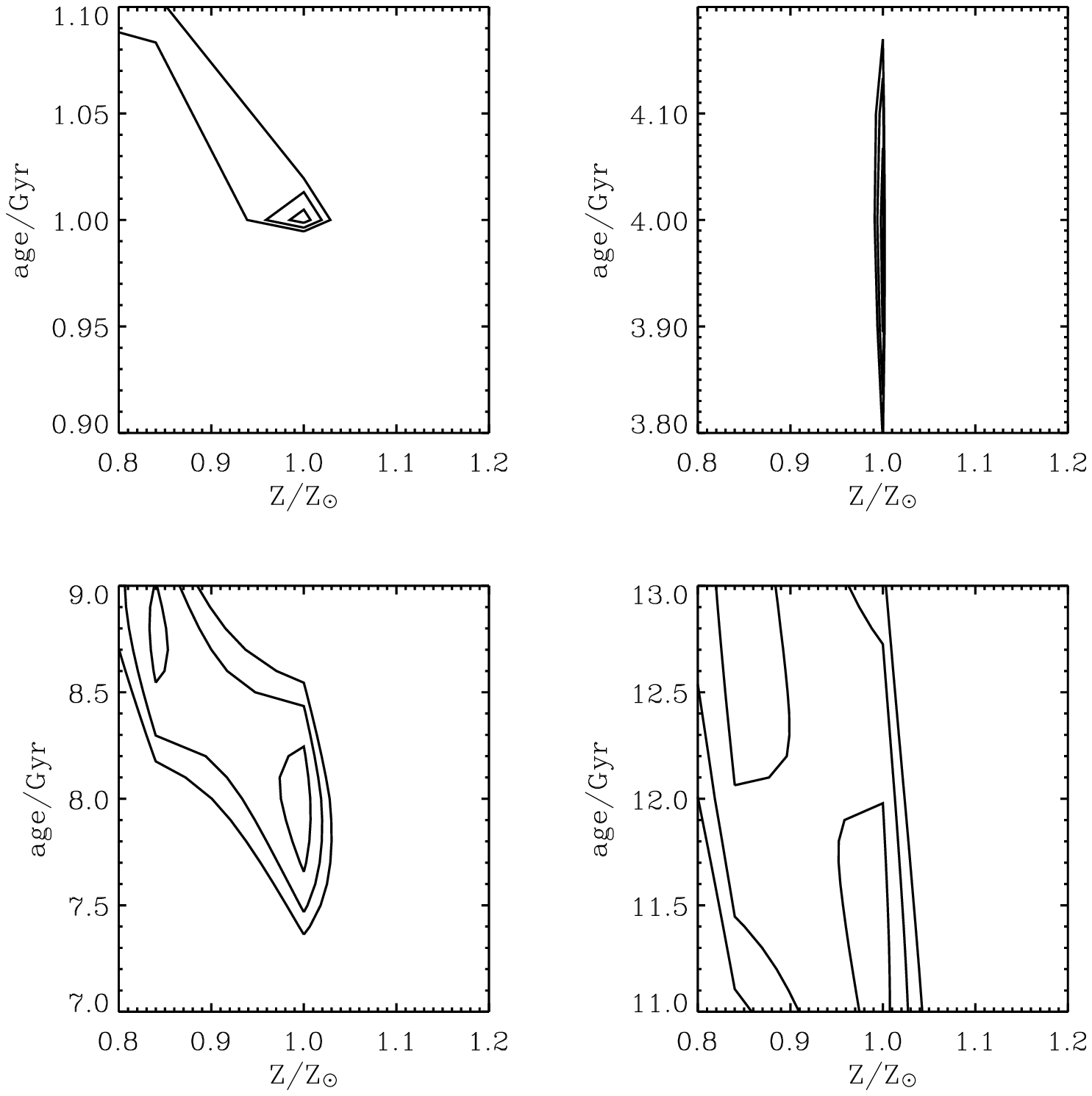}
\caption{Joint likelihood contours for age and metallicity. The four
panels correspond to spectra whose actual age is $1, 4, 8$ and $12$
Gyr and solar metallicity with spectral range from $4500$ to $7000$
\AA\, and with $S/N=10$ per 20\AA\,. The contours show the 68, 95 and
99\% confidence levels of recovering the true age and
metallicity. Note that there is some age-metallicity degeneracy,
although it is more pronounced for ages older than 8 Gyr.}
\label{fig:dege2}
\end{figure}

\begin{figure}
\includegraphics[width=8cm,height=8cm]{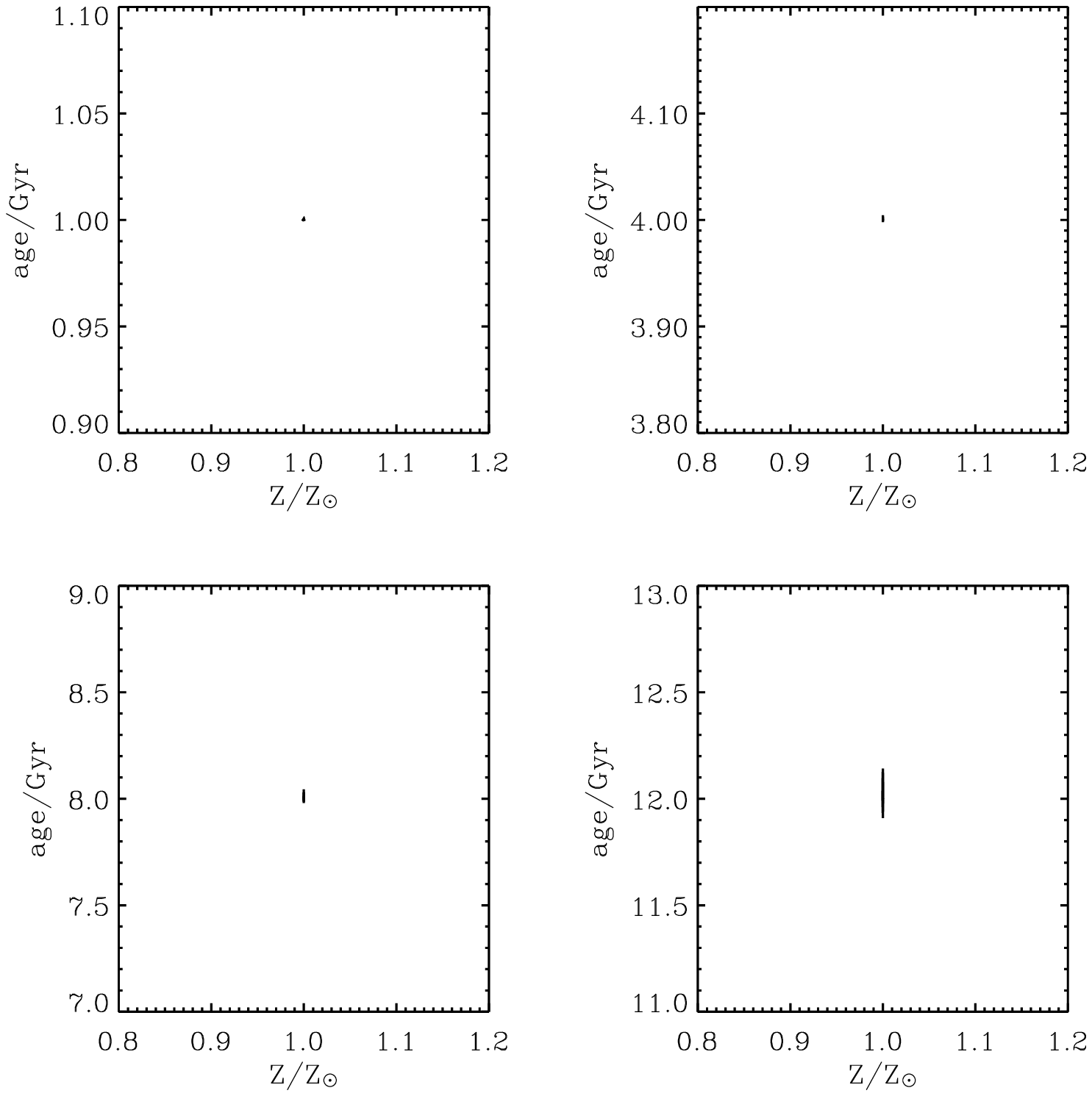}
\caption{Same as fig.~\ref{fig:dege2} but now the spectral range has
been increased to $2500-7000$ \AA\,. Note the total removal of the
age-metallicity degeneracy.} \label{fig:dege1}
\end{figure}

We define a stellar population as a collection of stars formed at a
certain rate in a certain chemical environment and whose integrated
light we are able to measure. We define a single stellar population
(SSP) as a stellar population formed in a uniform chemical environment
during a burst of infinitesimal duration (delta function). The steps
involved in building a SSP are:

\begin{enumerate}
\item A set of stellar tracks of different masses and of the metallicity of
the SSP is selected from our library.

\item An isochrone is built for the corresponding age. Each point in
the isochrone has a luminosity, effective temperature and gravity that
are used to assign the corresponding photospheric model.

\item The fluxes of all stars are summed up, with weights proportional to the
stellar initial mass function (IMF).
\end{enumerate}

SSPs are the building blocks of any arbitrarily complicated population
since the latter can be computed as a sum of SSPs, once the star
formation rate is provided. In other words, the luminosity of a
stellar population of age $t_0$ (since the beginning of star
formation) can be written as:

\begin{equation}
L_{\lambda}(t_0) = \int_0^{t_0} \int_{Z_i}^{Z_f} L_{\lambda}^{SSP} (Z, t_0-t) dZ dt
\end{equation}

where the luminosity of the SSP is:

\begin{equation}
L_{\lambda}^{SSP} (Z, t_0-t) = \int_{M_u}^{M_d} SF(Z,M,t)l_{\lambda}(Z, M, t_0-t) dM
\end{equation}

and $l_{\lambda} (Z, M, t_0-t)$ is the luminosity of a star of mass
$M$, metallicity $Z$ and age $t_0-t$, $Z_i$ and $Z_f$ are the initial
and final metallicities, $M_d$ and $M_u$ are the smallest and largest
stellar mass in the population and $SF(Z, M, t)$ is the star formation
rate at the time $t$ when the $SSP$ is formed. In this paper we
concentrate in building SSPs, while in previous papers we discuss
stellar populations that are consistent with chemical evolution
\citep{JPMH98,jim00}.

We have computed a large number of SSPs, with ages between $1 \times
10^6$ and $14 \times 10^9$ years, and metallicities from $Z=0.0002$ to
$Z=0.1$, that is from 0.01 to 5$Z_{\odot}$. Fig.~\ref{fig:fewssp}
shows a set of synthetic spectra of SSPs for three metallicities and
ages between $1 \times 10^7$ and $14 \times 10^9$ years (from top to
bottom).

\section{Parameter degeneracies: age and metallicity}

SSP are simple, given an IMF, they only depend on two parameters: age
and metallicity. It is therefore customary to ask the question: can
one recover these two parameters from an observed spectrum? Of course,
no single galaxy in the universe is a SSP, since its star formation
history is not going to be a single burst, nor its metallicity a
single one. However, most (if not all) elliptical galaxies have formed
their stellar populations at high redshifts in a very short duration
of time (e.g., \citet{JFDTPN99}) and they have relatively uniform chemical
compositions, thus their spectra can be correctly described
by a SSP.

The so called age-metallicity degeneracy has been the focus of
much attention during the past few years (e.g. \citet{W94}).
Several authors have argued that broad-band colors are degenerate
and therefore cannot be used to determine simultaneously age and
metallicity of a SSP. \citet{W94} has pointed out
that certain absorption features can be used to better determine
age and metallicity simultaneously. These feature are usually
confined to a relatively narrow spectral range ($4000-7000$ \AA).

In principle there is no need to discard information across the
spectrum since all of it may contain information that yields a joint
determination of the two parameters. Here we show how the whole
spectrum contains information about both age and metallicity and that
inclusion of light bluewards of $4500$ \AA\, helps to break any
degeneracy in the age-metallicity plane.

We first select a spectrum from the grid for a given age and
metallicity and add random noise. We then try to recover the age
and metallicity using $\chi^2$ minimization, marginalizing over
the amplitude. In the first case we limit the spectral range from
$4500$ to $7000$ \AA\ and add random noise with $S/N=10$ per
resolution element ($20$ \AA). We do this for four ages (1, 4, 8
and 12 Gyr) and solar metallicity. The corresponding contour plots
of the likelihood surfaces for 1,2 and 3$\sigma$ levels are shown
in fig.~\ref{fig:dege2}. Already, it can be seen that age and
metallicity are not completely degenerate since the contours do
close, however the {\em joint} determination of age and
metallicity is accurate only at the 20\% level. This also shows
that one does not need to concentrate in special spectral features
chosen empirically, the entire spectrum (despite the limited
spectral range considered) does contain information.

The above mild degeneracy is completely removed if we add 2000
\AA\, of bluer spectral light. Fig.~\ref{fig:dege1} shows similar
plots as the previous figure, but in this case the spectral range
of the spectrum goes from $2500$ to $7000$ \AA\, similar to that
of the SDSS survey. The figure shows that in this case age {\em
and} metallicity can be recovered with accuracy better than 1\%
(despite the moderate $S/N$ (10) of the mock test spectra).

\citet{DNJH03} reach similar conclusions using real spectra. They find
that addition of the same amount of UV light is sufficient to
completely determine the age-metallicity and star formation history of
an elliptical galaxy.

\begin{table}
\begin{center}
\begin{tabular}{|ccccc|}
\hline
Jimenez (true age/Gyr) & 1 & 4 & 8 & 12 \\
\hline
BC (recovered age/Gyr) & 1.2 & 3.5 & 7 & 11.1 \\
\hline
\end{tabular}
\caption{The age for a fiducial model created from our set of models by adding
  random noise with $S/N=10$ per resolution element of 20\AA\, as
  recovered by the Bruzual \& Charlot models. this gives an estimate
  of how important systematics are in determining the ages of stellar
  populations. From the above table it transpires that systematics in
  the absolute age are only at the 10\% level, while for the {\em
    relative} age are only at the $2-3$ \% level.}
\end{center}
\label{table:models}
\end{table}

\begin{table}
\begin{center}
\begin{tabular}{ccccc}
       & $\Delta (U-B)$ & $\Delta (B-V)$ & $\Delta (V-I)$ & $\Delta (B-K)$ \\
\hline
       &                &    0.5 Gyr     &                &                 \\
\hline
N-AGB  &   -0.10        &    -0.08       &   -0.10        &    -0.25       \\
R-AGB  &   +0.10        &    +0.08       &   +0.12        &    +0.30       \\
N-HB   &   -0.01        &    -0.05       &   -0.10        &    -0.20       \\
B-HB   &   -0.05        &    -0.02       &   -0.05        &    -0.05       \\
R-HB   &   +0.05        &    +0.02       &   +0.05        &    +0.04       \\
\hline
       &                &      5 Gyr     &                &                \\
\hline
N-AGB  &    0.00        &    0.00        &   -0.05        &    -0.10       \\
R-AGB  &    0.00        &    0.00        &   +0.05        &    +0.20       \\
N-HB   &   +0.01        &    0.00        &   -0.04        &    -0.20       \\
B-HB   &   -0.03        &    -0.02       &   -0.03        &    -0.05       \\
R-HB   &   0.00         &    0.00        &   +0.03        &    +0.03       \\
\hline
       &                &    14 Gyr      &                &                \\
\hline
N-AGB  &   +0.01        &   +0.01        &  +0.01         &   -0.08        \\
R-AGB  &   +0.05        &   +0.03        &  +0.10         &    +0.20       \\
N-HB   &   -0.10        &   -0.04        &  +0.04         &    +0.10       \\
B-HB   &   -0.15        &   -0.05        &  -0.05         &    -0.2        \\
R-HB   &   +0.05        &   -0.05        &  -0.05         &    -0.05       \\
\hline
\end{tabular}
\caption{The influence on the computed colours for a single synthetic
stellar population of changes in the assumptions underlying the
population computations. We list the changes corresponding to a
reference stellar population computed witha Salpeter IMF and solar
metallicity. N-AGB corresponds to models computed without an AGB,
while R-AGB corresponds to models computed with AGB but no mass
loss. N-HB stands for models computed with no HB. R-HB and B-HB are
models computed with red ($\eta=0.0$) and blue ($\eta=0.7$) HBs. The
different approaches can change the predicted colours by up to 0.3
magnitudes.}
\end{center}
\label{table:massloss}
\end{table}

\begin{figure*}
\includegraphics[width=14cm,height=14cm]{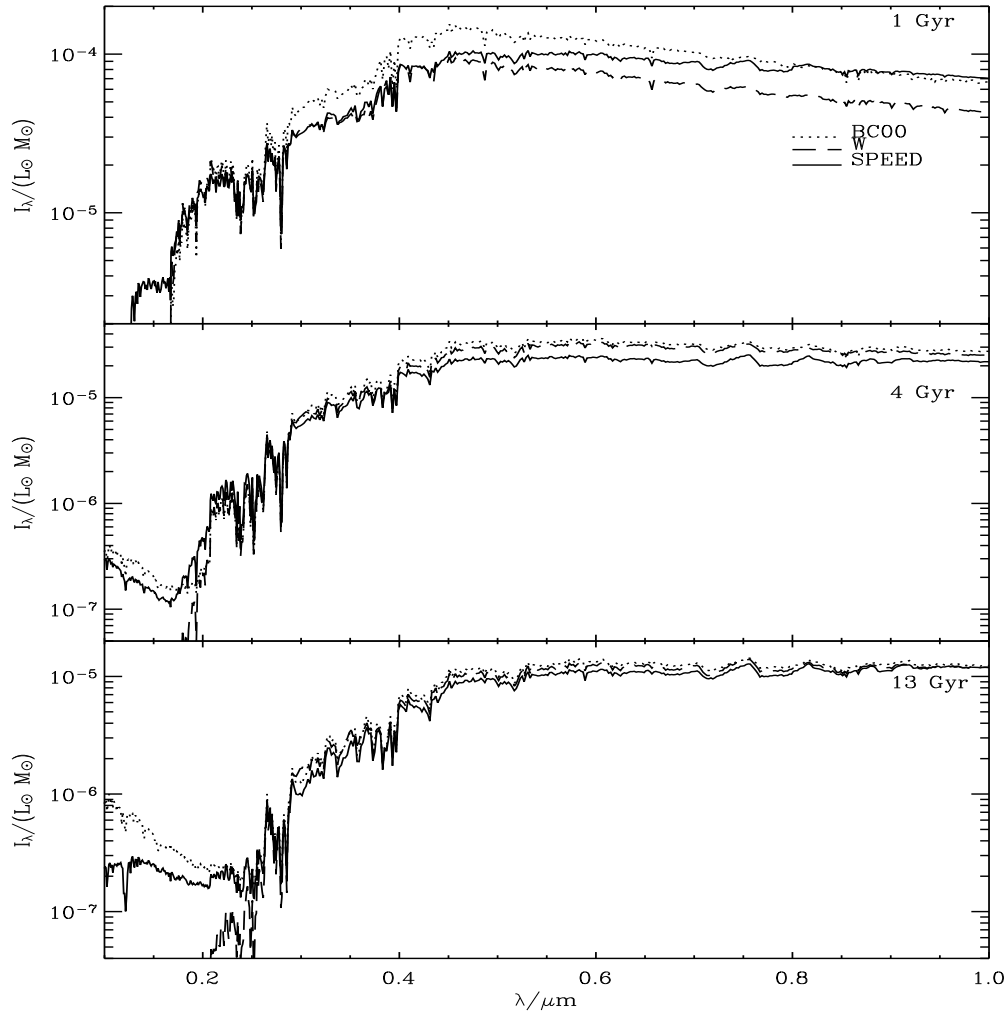}
\caption{Comparison between different single stellar population
models. The models are those by Bruzual \& Charlot (BC00), Worthey (W)
and ours (SPEED). The three models differ among each other, albeit
only at the $5-7\%$ level in flux. The discrepancy at wavelengths
smaller than 0.2 $\mu m$ between the models are due to different
treatment of post-AGB phases (which are not included in W94 models).}
\label{fig:comp_ssp}
\end{figure*}

\section{Systematic errors in SSPs}

In the previous section we have shown that given moderate $S/N$ and
wavelength spectral coverage, the parameters of a SSP can be
accurately recovered. What is the impact of systematics in this
parameter determination?

To answer this question and also to assess how our models compare to
previous work, we have done the following: first we have compared our
models to those by Bruzual \& Charlot (2000 models; \citet{BC93}) and
Worthey \citep{W94}. Fig.~\ref{fig:comp_ssp} shows the spectral
energy distributions as a function of wavelength for 3 different
ages. The 3 models differ among each other over the whole spectral
range. The models are more similar at older ages than at ages of 1
Gyr. The typical rms variation between our models and those of BC is
of $5-7$\% in intensity. How important is this difference? To address
this question we have performed the following test: we pick a spectrum
from our models (with a fixed age and metallicity) developed in this
paper and add gaussian noise with $S/N=10$ per resolution element of
20 \AA. We then attempt to recover the age using Bruzual \& Charlot
models. Results are presented in Table~\ref{table:models}. As can be
seen the systematic error in the {\em absolute} age is about 10\%
while the {\em relative} age difference is only of a $2-3$\%. We
therefore conclude that although systematic differences remain between
different models, their impact on parameter recovery is at the same
level as the difference in the intensity of the models.

We can also investigate the impact that mass loss can have in, for
example, the colours of integrated single stellar populations. In
Table~\ref{table:massloss} we show the change in the integrated
colours that different prescritions for mass loss, which translates
mostly in the morphology of the HB and termination of the AGB,
produce. The impact can be significant in the colors if mass loss is
ignored (as much as $0.3$ magnitudes in the infrared), thus careful
attention must be payed to this source of systematics.

\section{SSP at ultra-high redshift: Primordial Proto--Galaxies}

At $z>5$, traditional optical bands ($U, B, V, R$) fall below the rest
frame wavelength that corresponds to the Lyman break spectral feature
(1216 \AA), where most of the stellar radiation is extinguished either
by interstellar or intergalactic hydrogen. Because of this, galaxies
at $z>5$ are practically invisible at those photometric bands, and
even if they were detected, their colors would provide very little
information about their stellar population. New detectors and space
telescopes, such as SIRTF (http://sirtf.caltech.edu) , and in
particular the JWST (http://www.stsci.edu/jwst), will offer the
possibility of detecting very distant galaxies at IR wavelengths, and
of using photometric redshifts also with far--IR colors, as proposed
by \citet{SE99,Panagia+02}.

Nearby star forming galaxies are known to contain a considerable
amount of dust, that extinguishes a large fraction of their stellar UV
light and boosts, even by orders of magnitude, their IR
luminosity. The effect of dust must be taken into account in the
computation of colors involving photometric bands that span a large
wavelength interval. However, the search for very young
proto--galaxies, perhaps the first significant star forming systems in
the Universe, might be pursued without having to model the effects of
dust. The first stars to form in proto--galaxies must have primordial
chemical composition, or at least very low metallicity
(e.g. \citet{TSRBAP97,PJJ97}). Although it is difficult to model the
formation and disruption of dust grains on a galactic scale, it is
likely that the dust content of a galaxy grows together with its
metallicity, and that a proto--galaxy with primordial chemical
composition or very low metallicity ($Z\le 0.01 Z_{\odot}$) has
practically no dust (e.g. \citet{JPDBJM99}).  In this work we refer to
young protogalaxies with metallicity $Z\le 0.01 Z_{\odot}$ as
primordial protogalaxies, or PPGs. We show that, because of their very
blue colour, PPGs do not suffer from very strong cosmological dimming,
a feature which should aid their detectability in deep infrared
surveys with the JWST.

\subsection{Spectral energy distribution and IR colors of PPGs}

\begin{figure}
\includegraphics[width=8cm,height=8cm]{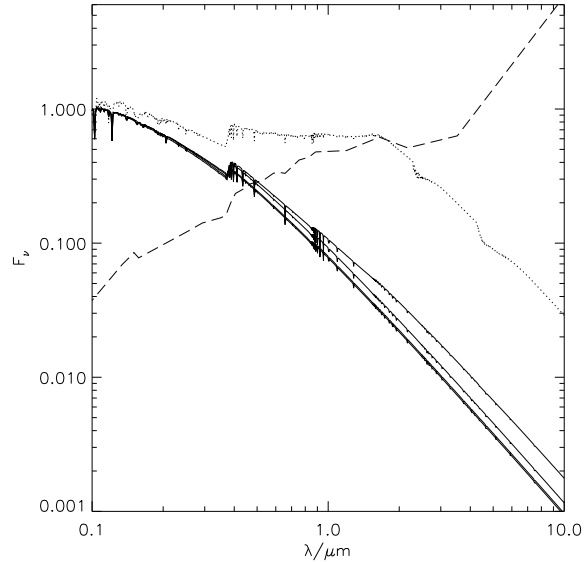}
\caption{ Spectral energy distribution for a starburst with constant star
  formation rate, age of 10 million years, metallicity $Z=0.01 Z_{\odot}$ and
  a Salpeter ($x=-1.35$) IMF with four different lower mass cutoffs: 0.1, 2.0,
  5.0 and 10.0 M$_{\odot}$ (solid lines), from top to bottom.  Starbursts with
  no low mass stars ($< 5$ M$_{\odot}$) are about a factor of 2 fainter than
  starbursts with low mass stars ($> 0.1$ M$_{\odot}$), at 3 $\mu$m. Note that
  for cutoff masses larger than 5 M$_{\odot}$, the spectral energy
  distribution changes very little, making our predictions almost independent
  of the exact value of the cutoff mass. The models have been normalised at
  1000 \AA. Also shown is a solar metallicity starburst from \citet{SKCS97}
  (dashed line). We finally plot an idealized starburst model with constant
  star formation rate, Salpeter IMF, age of 10 million years, $Z=0.2
  Z_{\odot}$ and no dust (dotted line). The UV SED of this starburst can be a
  rough approximation of the UV SED of quasars.}
\label{fig:IMF1}
\end{figure}

We have computed the SED of PPGs using the extensive set of synthetic
stellar population models developed in this paper. In particular, PPGs
are modelled using a constant star formation rate and $Z=0.01
Z_{\odot}$ metallicity. As noted above, we expect (and therefore
assume) that no photometrically significant quantity of dust has
already been formed in PPGs. We also assume that PPGs are young
objects (age less or equal than 100 million years). Furthermore, and
in order to test the shape of the primordial IMF, we have adopted a
Salpeter IMF ($x=-1.35$) with four different low mass cutoff values:
0.1, 2, 5 and 10 M$_{\odot}$ (i.e. the IMF does not contain any stars
with masses below the cut-off values).

The effect of the different low mass cutoff values on the SED of a PPG is
shown in fig.~\ref{fig:IMF1}, where the flux density F$_{\nu}$ is plotted in
arbitrary units for a 10 million year old PPG with $Z=0.01 Z_{\odot}$ and
constant star formation (solid lines).  Different solid lines corresponds to
the different IMF cutoffs, 0.1, 2, 5 and 10 M$_{\odot}$, from top to bottom
(the 5 and 10 M$_{\odot}$ are almost overlapped). As expected, the lack of low
mass stars translates into a deficit of flux at IR wavelengths. Furthermore,
for a low mass cutoff larger than 5 M$_{\odot}$ the SEDs do not differ much
since stars with masses larger than 5 M$_{\odot}$ have similar spectral energy
distributions in the IR. The difference in relative flux density over the wide
wavelength range illustrated in fig.~\ref{fig:IMF1} provides the means to
test the IMF in PPGs.  The SED of an observed nearby starburst, from
\citet{SKCS97}, is also plotted in fig.~\ref{fig:IMF1} (dashed line),
together with a 10 million year model starburst, with continuous star
formation, metallicity $Z=0.2 Z_{\odot}$, and no dust extinction or emission
(dotted line).  The difference in the spectral slope between PPGs and nearby
starbursts is striking, and it is still significant between PPGs and the
idealized starburst model with no dust. Although the idealized starburst model
with no dust is unlikely to describe any galaxy observed nearby, it can be
used as an approximate model for the UV SED of quasars, due to its rather flat
SED in the far ultra violet wavelengths.

\begin{figure}
\includegraphics[width=9cm,height=8cm]{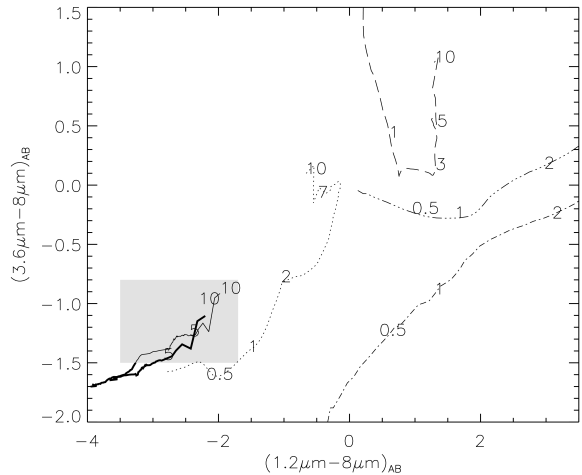}
\caption{Colour--colour (in the AB system) redshift evolution of:
the observed starburst of solar metallicity from Figure~11 (dashed
line); the idealized starburst model of 1/5 solar metallicity and
no dust, also from Figure~11 (dotted line); two PPGs models (from
Figure~11), with constant star formation rate, age of 0.01~Gyr,
$Z=0.01 Z_{\odot}$, and Salpeter IMF with 0.1~M$_{\odot}$ cutoff
(thin solid line) and 5~M$_{\odot}$ cutoff (thick solid line); two
typical nearby galaxies (a spiral and an elliptical), from
\citet{SKCS97}. The shaded area shows where PPGs are expected to
be.} \label{fig:IMF2}
\end{figure}

We have computed AB magnitudes ($m_{\rm AB}=-2.5{\rm log}(F_{\nu})-48.59$,
where $F_{\nu}$ is expressed in erg s$^{-1}$ Hz$^{-1}$) for 3 different
wavelengths: 1.2, 3.6 and 8 $\mu$m. Fig.~\ref{fig:IMF2} shows the
colour--colour trajectory for PPGs with a Salpeter IMF with a low mass cutoff
of 0.1 M$_{\odot}$ (thin solid line), and 5 M$_{\odot}$ (thick solid line),
metallicity $Z=0.01 Z_{\odot}$, and an age of 10 million years.  The dashed
line is the evolution of the nearby starburst from fig.~\ref{fig:IMF2}, and
the dotted line the evolution of the idealized starburst model with no dust
and $Z=0.2Z_{\odot}$, also from fig.~\ref{fig:IMF2}.  All models have been
simply K--corrected and the numbers that label the trajectories indicate the
corresponding redshift. The figure shows that PPGs with redshift range
$5<z<10$ have colors in the range of values
$-3.0\le(1.2\mu$m$-8\mu$m$)_{AB}\le-2.0$ and
$-0.9\le(3.6\mu$m$-8\mu$m$)_{AB}\le-1.5$, that are inside the grey area. The
idealized starburst model with no dust and sub--solar metallicity enters
marginally the dashed area, but only for low redshifts ($z<0.5$). This
idealized model is unlikely to describe nearby galaxies, while it can be a
rough description of the UV SED of quasars, in which case it could be
concluded that PPGs should have very different colors than quasars at redshift
$z>0.5$ (at least about 0.5~mag away in the color--color plot of
fig.~\ref{fig:IMF2}).  Nearby galaxies are either star forming galaxies with
significant amount of dust (irregular, spiral or starburst galaxies), or older
stellar systems with little gas or dust (elliptical galaxies). In both cases
nearby galaxies are much redder than PPGs. In fig.~\ref{fig:IMF2}, the
dashed dotted lines show the color--color redshift evolution of two typical
nearby galaxies (a spiral and an elliptical galaxies, from \citet{SKCS97}).
These galaxies, even if nearby, are always at least 2~mag redder than PPGs in
the $(1.2\mu$m$-8\mu$m$)_{AB}$ color.

The fact that PPGs are the bluest objects, in this IR color--color plot, is an
important result, since a photometric search for high redshift galaxies would,
in principle, be biased toward selecting the solar metallicity starbursts,
which are the reddest galaxies, as already proposed by \citet{SE99}. Although
it cannot be excluded that galaxies of metallicity close to the solar value
might exist at $z=10$, and that PPGs are rare (since they are young by
definition), fig.~\ref{fig:IMF2} shows that primordial star formation at
high redshift should be searched for in very blue objects. The shaded area in
fig.~\ref{fig:IMF2} marks the expected location of PPGs.

\section{The IR Flux Density of PPGs}

\begin{figure}
  \includegraphics[width=9cm,height=8cm]{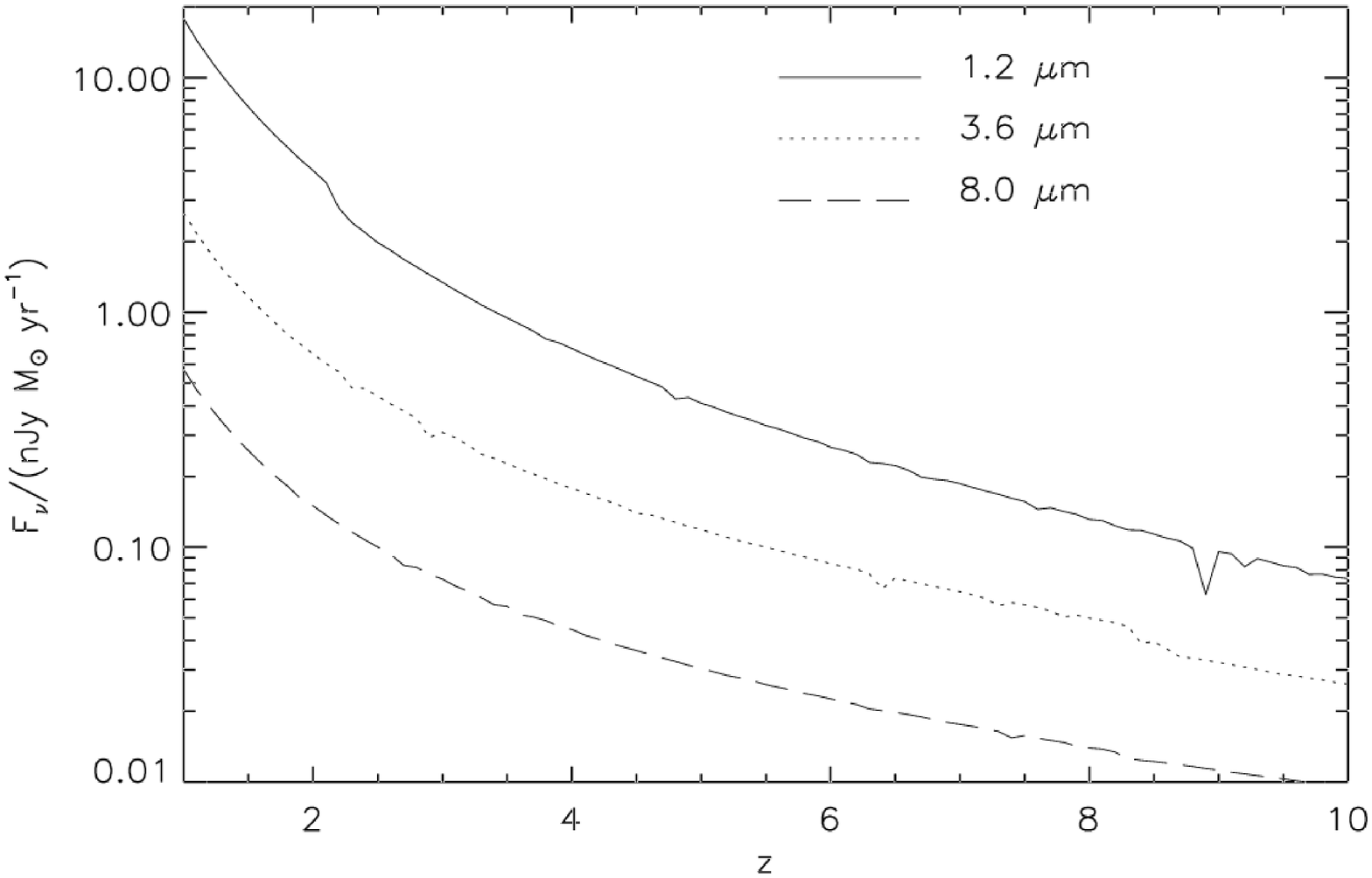}
  \caption{The expected IR flux density of PPGs, per unit of star
  formation rate, as a function of redshift, for four wavelengths and
  2 cosmologies (in both cases $H_0=65$ km s$^-1$ Mpc$^{-1}$). Note
  that the cosmological dimming is very small since PPGs have $F_{\nu}
  \propto \nu^{1.5}$ (see Figure~11). The same model as in Figures~11
  and 12 has been used, with a 5~M$_{\odot}$ cutoff IMF.  The flux
  density has been computed for an age of 0.01~Gyr, and increases
  rather slowly afterward.}
\label{fig:IMF3}
\end{figure}

We have just shown that PPGs can be photometrically selected as
the bluest galaxies in the Universe.  The question that needs to
be answered now is: Can PPGs be detected at all with future
telescopes such as the SIRTF and the JWST? PPGs could in fact be
very faint because they could be very small, or because their star
formation rate (SFR) could be very low. To answer this question,
we have computed the expected flux in nJy per unit of SFR (in
M$_{\odot}/$yr), as a function of redshift, for a PPG with a
Salpeter IMF and cutoff mass of 5 M$_{\odot}$, in a $\Lambda$
dominated Universe ($\Omega_m=0.3$, $\Lambda=0.7$, $H_0=65$ km
s$^{-1}$ Mpc$^{-1}$). The plots are roughly independent of the
starburst age, for age larger than a few million years.
Fig.~\ref{fig:IMF3} shows that, although PPGs do not exhibit a
negative K-correction as galaxies do in sub-mm and mm bands, they
suffer from very little cosmological dimming, even in an open
Universe. From fig.~\ref{fig:IMF3}, it can be seen that PPGs with
SFR of about $100$ M$_{\odot}/$yr have a flux in the faintest band
($8\mu$m) of about 1 nJy (although they would be much brighter at
1.2 $\mu$m -- about 10-20 nJy). The IRAC camera at SIRTF will be
able to measure in 1 hour exposure a flux of 500nJy at the
5$\sigma$ level, and will not be able to detect PPGs which a SFR
of less than 100 $M_{\odot}$ yr$^{-1}$ but much higher. On the
other hand, the JWST time calculator (htpp:/www.stsci.edu/jwst)
gives (for a point source using the MIR-ACCUM instrument with a
resolution of 3, S/N=5) that exposing 300 hours one can get a flux
of about 5nJy at 8$\mu$m. In order to distinguish PPGs we also
need observations at 3.6 and 1.2 $\mu$m (see fig.~\ref{fig:IMF2}).
using the same calculator we obtain (this time NIR-ACCUM,
resolution 3, S/N=5) that it takes 3.63 hours to achieve 1 nJy at
3.6 $\mu$m, while it takes 3.12 hours at 1.2 $\mu$m. It is worth
noting though that PPGs are expected to be about 1 order of
magnitude brighter at 1.2 and 3.6 than at 8 $\mu$m. Therefore, for
the 4 nJy constraint imposed by the 8 $\mu$m band one would need
only a few minutes to achieve the required sensitivity at 1.2 and
3.6 $\mu$m. It is therefore accurate to say that the limit on our
observations comes from the 8 $\mu$m band solely. Now, inspection
of fig.~\ref{fig:IMF3} shows that a flux of 5nJy at 8$\mu$m is
achieved for PPGs with SFRs of about 400 M$_{\odot}$ yr$^{-1}$.
For example, one needs to have systems with ages of 10$^7$ yr and
masses of a few $10^9$ M$_{\odot}$ in order for them to achieve a
flux of 4nJy. This is not an unreasonable mass for the first
objects since it is likely that star formation of the first
objects will be retarded until a relatively large halo is in place
\citep{OhHaiman03}.

\subsection{The Primordial Stellar IMF}

Observational and theoretical arguments suggest that stars forming
from gas of very low metallicity ($ < 10^{-3} Z_{\odot}$) could have
an initial mass function (IMF) shifted toward much larger masses than
stars formed later on from chemically enriched gas.  The observational
arguments are extensively discussed in a paper by \citet{L98}, and we
refer the reader to that work. The main theoretical reason in favor of
a `massive' zero metallicity IMF is the relatively high temperature
($T>100$~K) of gas of primordial composition, that is cooled at the
lowest temperatures mainly by H$_2$ molecules
\citep{PSS83,MS86,SK87,KSFR90,KS92,AN96}. One of the most exciting
aspects of the photometric discovery of PPGs would be the possibility
of investigating the nature of their stellar IMF, once their redshifts
are available.

It is likely that while
the local properties of the ISM do not interfere significantly with the self
similar dynamics that originate the stellar IMF, they do play a role in
setting the particular value of the mass scale where the self similarity is
broken.  According to this point of view, one expects the stellar IMF to have
always more or less the same power law shape, down to a cutoff mass whose
value depends on local properties of the ISM, and up to the largest stellar
mass, whose value is limited either by the total mass of the star formation
site, or by some physical process that prevents the formation of
super--massive stars.

The lower mass cutoff of the stellar IMF has been predicted in models of i)
opacity limited gravitational fragmentation \citep{S177,S277,S377,YS85,YS86};
ii) protostellar winds that would stop the mass accretion onto the proto-star
\citep{AF96}; iii) fractal mass distribution with fragmentation down to one
Jeans' mass \citep{L92}.  If gravitational sub--fragmentation during collapse
is not very efficient (see \citet{B93}), the value of the Jeans' mass
determines the lower mass cutoff of the IMF. In \citet{PNJ97}, numerical
simulations of super--sonic and super--Alfv\'{e}nic \citep{PN99}
magneto--hydrodynamic turbulence are used to compute the probability density
function of the gas density, which is used to predict the distribution of the
Jeans' mass in turbulent gas, under the reasonable assumption of uniform
kinetic temperature.  The Jeans' mass distribution computed in
\citet{PNJ97} has an exponential cutoff below a certain mass value, that
is found to be:
\begin{equation}
M_{\rm min}= 0.2M_{\odot}\left(\frac{n}{10^3cm^{-3}}\right)^{-1/2}
\left(\frac{T}{10 K}\right)^{2}\left(\frac{\sigma_{v}}{2.5 km/s}\right)^{-1}
\label{1}
\end{equation}
where $T$ is the gas temperature, $n$ the gas density, and $\sigma_{v}$ the
gas velocity dispersion. Using the ISM scaling laws, according to which
$n^{1/2}\sigma_{v}\approx const$, one obtains:
\begin{equation}
M_{\rm min}\approx 0.1 M_{\odot}\left(\frac{T}{10K}\right)^2
\label{2}
\end{equation}
that is a few times smaller that the average Jeans mass (the
Jeans mass corresponding to the average gas density), and therefore an
important correction to more simple models of gravitational fragmentation,
that do not take into account the effect of super--sonic turbulence on the gas
density distribution.

If the ISM has primordial chemical composition, and the main coolant is
molecular hydrogen, a temperature below $100$~K is hardly reached, and the
stellar IMF might have a lower mass cutoff of about $10$~M$_{\odot}$. Similar
lower mass cutoffs are obtained in the models by \citet{S377} and
\citet{YS86}, who estimated typical stellar masses, based on molecular
hydrogen cooling, of approximately $20$ and $10$~M$_{\odot}$ respectively.
More recent numerical simulations of the collapse and cooling of cosmological
density fluctuations of large amplitude (the first objects to collapse in the
Universe), yield even larger values of the Jeans' mass, of the order of
$100$~M$_{\odot}$ (Bromm, Coppi, \& Larson 1999; Abel, Bryan, \& Norman 1998)

\begin{figure}
\includegraphics[width=8cm,height=8cm]{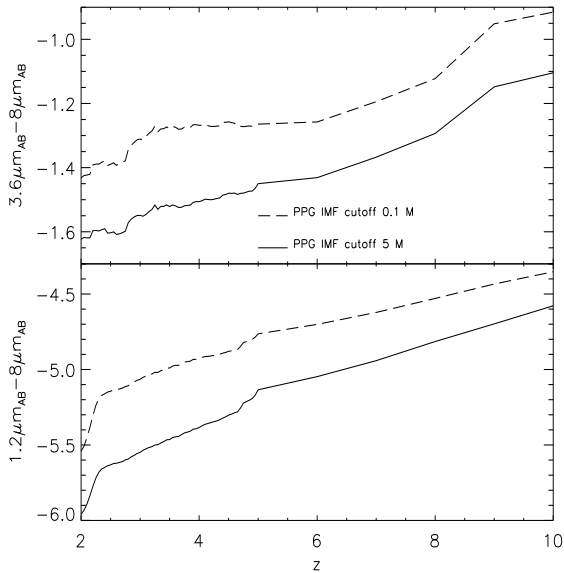}
\caption{Colour evolution (AB system) with redshift for PPGs with a
  Salpeter IMF with $0.1$~M$_{\odot}$ cutoff (dashed line), and
  $5$~M$_{\odot}$ cutoff (solid line). As expected from Figure~11, the biggest
  difference occurs for colours that sample both the near-IR (rest--frame UV)
  and the far-IR. The difference between the Salpeter and the 'massive' IMF is
  about 0.5 mag in $(1.2\mu$m$-8\mu$m$)_{AB}$. Once PPGs have been found at
  high $z$ using the broad band color--color selection from Figure~12, and
  their redshift has been determined with narrow band photometry, the
  $(1.2\mu$m$-8\mu$m$)_{AB}$ color can be used to discriminate among objects
  with a standard Salpeter IMF or a $5$~M$_{\odot}$ cutoff IMF.}
\label{fig:IMF4}
\end{figure}

The discovery of PPGs could shed new light on the problem of the primordial
IMF. The redshift evolution of the two colors $(1.2\mu$m$-8\mu$m$)_{AB}$ and
$(3.6\mu$m$-8\mu$m$)_{AB}$, computed with the PPG model discussed in this
work, is plotted in fig.~\ref{fig:IMF4}. The solid line is the case of a PPG
with a Salpeter IMF with $5$~M$_{\odot}$ cutoff, and the dashed line the same
PPG model, but with a $0.1$~M$_{\odot}$ cutoff. Once a PPG candidate is
selected with the IR broad band photometry as a very blue object (colors
inside the shaded area in fig~\ref{fig:IMF2}), and its redshift is
estimated with a Lyman drop method, or with an H$\alpha$ search, the IR
colours provide a tool to discriminate between a standard IMF, and an IMF
deprived of low mass stars.  The lower panel of fig~\ref{fig:IMF4} shows
that a PPG with a 'massive' IMF can be about 0.5~mag bluer in
$(1.2\mu$m$-8\mu$m$)_{AB}$ than a PPG with a standard IMF.

\subsection{Detectability via broad-band infrared imaging versus
  emission-line searches} In this work we propose to select PPGs as the bluest
objects in deep IR surveys, on the basis of the color--color plot
shown in fig.~2. We now address the question of how a broad band
photometric selection of PPGs performs, compared with searches of
emission lines, such as Lyman-$\alpha$ and H$\alpha$.  The rest
frame equivalent widths of Lyman-$\alpha$ and H$\alpha$ can be
very roughly estimated by assuming that each photon below 1251 and
1025 \AA\, will originate a Lyman-$\alpha$ and H$\alpha$ photon
respectively.  The equivalent widths estimated in this way are of
course upper limit to the true equivalent widths.  We find that
the equivalent width of Lyman-$\alpha$ is 380 \AA\, while the
equivalent width of H$-\alpha$ is 4400 \AA\ --the latter is so
high due to the fact that the continuum at 6563 \AA\ is rather
faint in PPGs (see Fig.~\ref{fig:IMF1}). Assuming that the lines
have intrinsic widths at rest--frame typical of the virial
velocity of a galaxy (for example a line width of 300~km/s
corresponds to 2 and 13 \AA\ respectively), one finds that they
will only be about 30 times brighter than the continuum at $z \sim
10$. Since one would need to shift the narrow filter for about 500
steps, or more, to search for all possible emitters in the
redshift range $5\le z\le 10$, the advantage of the lines being
brighter than the continuum is offset by the number of steps
needed to find all PPGs between $z=5$ and 10. It seems therefore
that IR broad band photometry is an easier way to both detect and
select PPG candidates than the emission line technique, because
only one deep exposure is needed to find {\it all} PPGs in the
redshift range $5\le z\le 10$. Note that the equivalent width of
Lyman-$\alpha$ and H$\alpha$ has been over--estimated here.
Moreover, the advantage of deep broad band photometry is that,
together with detecting and selecting PPGs, it provides at the
same time important information about their stellar populations.
However, it is important that the emission line technique (or a
Lyman break technique) is available aboard the NGST, since
photometric redshifts measured with narrow filters will probably
be the best (or the only) way to further constrain the redshift of
PPG broad band photometric candidates, which is necessary to
extract information about their stellar population from the broad
band colors (fig.~4)\footnote{We note in passing that the
acquisition of photometry
  around the Lyman break to determine their redshift would not be time
  consuming as shown before and that the biggest limitation in detecting PPGs
  is due to the poor performance of the 8 $\mu$m detectors.}.

A star formation rate of $100$~M$_{\odot}/$yr over a few million years
is necessary for detecting a PPG with the JWST. With such SFR, after
$10^7$ years $10^9$~M$_{\odot}$ of gas is turned into stars. In order
to still have a metallicity of $Z\le 0.01 Z_{\odot}$, these stars must
be formed in a system with baryonic mass of at least $1\times
10^{11}$~M$_{\odot}$, that is a very large galaxy, or a small group of
galaxies. Such massive systems are inside large dark matter halos that
are not collapsing yet at redshift $5\le z\le 10$. It is possible that
PPGs that can be detected with the JWST are the progenitors of very
large galaxies, in a phase when their dark matter halo has not turned
around yet.  If PPGs are discovered, their spectro--photometric
properties could give very important clues for the problem of star
formation in galaxies (such as the origin of the stellar IMF) and
their luminosity, abundance, and redshift distribution would trace the
complete history of the very first star formation sites in the
Universe.

\section{Conclusions}

In this paper we have presented new stellar interior tracks and single stellar
population models of arbitrary age, metallicity and IMF. Our main findings are:

\begin{enumerate}

\item A new set of stellar interior models has been presented. The
models cover {\em all} stages of stellar evolution and we have built
isochrones out of them for ages between $10^6 - 1.4 \times 10^{10}$.

\item It is possible to construct stellar evolution models that
accurately reproduce the properties of individual stars for a wide
range of ages and metallicities.

\item We have presented a new algorithm to compute the evolution of stars in
  the RGB, HB and AGB. This algorithm makes it possible to explore the effect
  of variations in some of unknown parameters of stellar evolution like mass
  loss, mixing length etc.

\item We have developed a new and fast algorithm to build synthetic stellar
  evolution spectra and colour--magnitude diagrams of arbitrary metallicity
  and age.

\item We have shown that changes in the values of the stellar parameters like
  mass loss and mixing length can change the predicted colours of a population
  by as much as 0.4 mag.

\item We have studied degeneracies in the parameter space (age and
metallicity) and shown that these parameters are only degenerated if
the wavelength range of the spectrum is very small or only a few
spectral features are chosen. Addition of light bluewards of the
$4500$ \AA\, ($2500 - 4500$ \AA) significantly reduces this
degeneracies and, in fact, lifts them.

\item It has been shown that systematic errors among different models are at the
level of 10-20\%, despite using completely different stellar input physics. It should
be possible to reduce these errors even further.

\item We have studied the photometric properties of very young proto--galaxies
  with primordial or very low ($Z=0.01 Z_{\odot}$) metallicity and no
  significant effect of dust in their SED. We have named these galaxies
  ``primordial protogalaxies'', or PPGs. Using the methods of synthetic
  stellar populations, we predict that PPGs are the bluest stellar systems
  in the Universe. They can therefore be selected in color--color diagrams
  obtained with deep broad band IR surveys, and can be detected with the JWST,
  if they have a SFR of at least $100$~M$_{\odot}/$yr, over a few million
  years.  We have discussed the possibility of using the IR colours of PPGs to
  constrain their stellar IMF, and investigated the possibility that the
  stellar IMF arising from gas of primordial chemical composition is more
  ``massive'' than the standard Salpeter IMF. Finally we have argued that
  broad band photometry can be more efficient than emission line searches, to
  detect and select PPGs.

\end{enumerate}

The models are available on the world wide web ({\tt
www.roe.ac.uk/$\sim$jsd} and {\tt www.physics.upenn.edu/$\sim$raulj}). We
provide the stellar interior tracks presented in this paper and
the single stellar population models and tools to compute
photometry and synthetic stellar populations with arbitrary star
formation histories.

\section*{acknowledgments}
We thank the referee, Guy Worthey, for comments that greatly
improved this paper. RJ thanks Eric Agol and Marc Kamionkowski for
encouraging him to publish the stellar models contained in this
paper. The work of RJ is partially supported by NSF grant
AST-0206031. James Dunlop acknowledges the enhanced research time
provided by the award of a PPARC Senior Fellowship.

\clearpage
\section{Appendix: Analytic fits for colours of SSPs}

The magnitudes for a SSP (normalized to 1 M$_{\odot}$) as a function of age
and metallicity, for a given photometric band UBVRIJK, are approximated to
within 4\% by:
\begin{equation}
M_{\lambda}=-2.5 \times \sum_{i=0}^{4}\sum_{j=0}^{4} X^i C_{\lambda}(i+1,j+1) Y^j,
\end{equation}
where
\begin{eqnarray}
X & = & 5.76+3.18  \log{\tau}+1.26  \log^2{\tau}+2.64
\log^3{\tau} \nonumber \\
& & +1.81  \log^4{\tau} +0.38  \log^5{\tau},
\end{eqnarray}
\begin{eqnarray}
Y & = & 2.0+2.059  \log{\zeta}+1.041  \log^2{\zeta} \nonumber \\
& & +0.172 \log^3{\zeta} - 0.042  \log^4{\zeta},
\end{eqnarray}
and
\begin{equation}
\tau=\frac{t}{\rm Gyr}, \;\;\;\;\;\;\;\;\;\;\;\; \zeta = \frac{Z}{Z_\odot}.
\end{equation}

Luminosities are obtained simply from $L_{\lambda}=10^{-0.4*(M_{\odot
\lambda}- M_{\lambda})}$, where $M_{\odot \lambda}=\{5.61, 5.48, 4.83, 4.34,
4.13, 3.72, 3.36, 3.30, 3.28\}$ for $\{U, B, V, R, I, J, H, K, L\}$.  The $i$
and $j$ values appear as exponents of $X$ and $Y$, respectively, and as
indexes defining elements of the $C_\lambda$ matrices, given by

\small
\begin{equation}
C_U=\left( \begin{array}{rrrrr}
  -4.738\times 10^{-1}&  4.029\times 10^{-1}& -3.690\times 10^{-1}&  1.175\times 10^{-1}& -1.253\times 10^{-2}\\
  -2.096\times 10^{-1}& -1.743\times 10^{-1}&  1.268\times 10^{-1}& -2.526\times 10^{-2}&  8.922\times 10^{-4}\\
  -1.939\times 10^{-2}&  1.401\times 10^{-2}& -9.628\times 10^{-3}& -1.754\times 10^{-3}&  7.237\times 10^{-4}\\
   2.671\times 10^{-3}& -4.271\times 10^{-4}&  2.470\times 10^{-4}&  2.963\times 10^{-4}& -7.334\times 10^{-5}\\
  -7.468\times 10^{-5}&  6.676\times 10^{-7}&  1.482\times 10^{-6}& -9.594\times 10^{-6}&  2.055\times 10^{-6}
\end{array} \right);
\end{equation}

\begin{equation}
C_B=\left( \begin{array}{rrrrr}
  -8.321\times 10^{-1}&  5.972\times 10^{-1}& -4.818\times 10^{-1}&  1.356\times 10^{-1}& -1.271\times 10^{-2}\\
  -1.223\times 10^{-1}& -2.523\times 10^{-1}&  2.117\times 10^{-1}& -5.309\times 10^{-2}&  3.716\times 10^{-3}\\
  -2.632\times 10^{-2}&  2.468\times 10^{-2}& -2.462\times 10^{-2}&  4.163\times 10^{-3}&  5.054\times 10^{-5}\\
   2.835\times 10^{-3}& -7.906\times 10^{-4}&  9.653\times 10^{-4}& -1.164\times 10^{-5}& -3.800\times 10^{-5}\\
  -7.416\times 10^{-5}&  1.120\times 10^{-6}& -6.707\times 10^{-6}& -5.615\times 10^{-6}&  1.611\times 10^{-6}
\end{array} \right);
\end{equation}
\begin{equation}
C_V=\left( \begin{array}{rrrrr}
  -9.348\times 10^{-1}&  7.376\times 10^{-1}& -5.170\times 10^{-1}&  1.182\times 10^{-1}& -8.023\times 10^{-3}\\
  -9.521\times 10^{-2}& -3.476\times 10^{-1}&  2.379\times 10^{-1}& -4.485\times 10^{-2}&  1.303\times 10^{-3}\\
  -2.437\times 10^{-2}&  4.413\times 10^{-2}& -2.802\times 10^{-2}&  2.202\times 10^{-3}&  5.271\times 10^{-4}\\
   2.625\times 10^{-3}& -2.063\times 10^{-3}&  9.257\times 10^{-4}&  2.378\times 10^{-4}& -8.471\times 10^{-5}\\
  -6.994\times 10^{-5}&  2.842\times 10^{-5}&  7.436\times 10^{-7}& -1.424\times 10^{-5}&  3.050\times 10^{-6}
\end{array} \right);
\end{equation}
\begin{equation}
C_R=\left( \begin{array}{rrrrr}
  -9.755\times 10^{-1}&  8.121\times 10^{-1}& -4.535\times 10^{-1}&  5.553\times 10^{-2}&  3.313\times 10^{-3}\\
  -7.346\times 10^{-2}& -4.000\times 10^{-1}&  1.983\times 10^{-1}& -6.018\times 10^{-3}& -5.621\times 10^{-3}\\
  -2.368\times 10^{-2}&  5.415\times 10^{-2}& -1.968\times 10^{-2}& -5.209\times 10^{-3}&  1.811\times 10^{-3}\\
   2.502\times 10^{-3}& -2.667\times 10^{-3}&  1.376\times 10^{-4}&  8.392\times 10^{-4}& -1.845\times 10^{-4}\\
  -6.690\times 10^{-5}&  4.003\times 10^{-5}&  2.413\times 10^{-5}& -3.043\times 10^{-5}&  5.650\times 10^{-6}
\end{array} \right);
\end{equation}
\begin{equation}
C_I=\left( \begin{array}{rrrrr}
  -1.027\times 10^{0} &  8.951\times 10^{-1}& -3.759\times 10^{-1}& -1.908\times 10^{-2}&  1.690\times 10^{-2}\\
  -4.672\times 10^{-2}& -4.514\times 10^{-1}&  1.411\times 10^{-1}&  4.385\times 10^{-2}& -1.440\times 10^{-2}\\
  -2.379\times 10^{-2}&  6.399\times 10^{-2}& -8.885\times 10^{-3}& -1.425\times 10^{-2}&  3.380\times 10^{-3}\\
   2.414\times 10^{-3}& -3.267\times 10^{-3}& -7.741\times 10^{-4}&  1.532\times 10^{-3}& -3.012\times 10^{-4}\\
  -6.417\times 10^{-5}&  5.166\times 10^{-5}&  4.964\times 10^{-5}& -4.843\times 10^{-5}&  8.607\times 10^{-6}
\end{array} \right);
\end{equation}
\begin{equation}
C_J=\left( \begin{array}{rrrrr}
  -1.106\times 10^{0} &  1.043\times 10^{0} & -1.932\times 10^{-1}& -1.715\times 10^{-1}&  4.364\times 10^{-2}\\
   1.122\times 10^{-2}& -5.240\times 10^{-1}& -6.001\times 10^{-3}&  1.510\times 10^{-1}& -3.228\times 10^{-2}\\
  -2.757\times 10^{-2}&  7.613\times 10^{-2}&  1.935\times 10^{-2}& -3.358\times 10^{-2}&  6.540\times 10^{-3}\\
   2.502\times 10^{-3}& -3.923\times 10^{-3}& -2.956\times 10^{-3}&  2.944\times 10^{-3}& -5.274\times 10^{-4}\\
  -6.439\times 10^{-5}&  6.234\times 10^{-5}&  1.066\times 10^{-4}& -8.371\times 10^{-5}&  1.417\times 10^{-5}
\end{array} \right);
\end{equation}
\begin{equation}
C_K=\left( \begin{array}{rrrrr}
  -1.132\times 10^{0} &  1.296\times 10^{0} & -1.795\times 10^{-1}& -2.391\times 10^{-1}&  5.794\times 10^{-2}\\
   6.838\times 10^{-2}& -6.627\times 10^{-1}& -5.987\times 10^{-2}&  2.115\times 10^{-1}& -4.319\times 10^{-2}\\
  -3.194\times 10^{-2}&  1.009\times 10^{-1}&  3.092\times 10^{-2}& -4.459\times 10^{-2}&  8.454\times 10^{-3}\\
   2.649\times 10^{-3}& -5.499\times 10^{-3}& -3.923\times 10^{-3}&  3.748\times 10^{-3}& -6.625\times 10^{-4}\\
  -6.607\times 10^{-5}&  9.556\times 10^{-5}&  1.332\times 10^{-4}& -1.038\times 10^{-4}&  1.746\times 10^{-5}
\end{array} \right).
\end{equation}
\normalsize

\end{document}